\documentstyle[aps,manuscript]{revtex}

\begin{document}
\draft
\preprint{NSF-ITP 94-104}
\title{Non-Ergodic Dynamics of the 2D Random-Phase Sine-Gordon Model:
Applications to Vortex-Glass Arrays and Disordered-Substrate Surfaces}
\author{D. Cule and Y. Shapir }
\address{ Institute for Theoretical Physics\\University of California\\
Santa Barbara, CA 93106-4030\\and\\ Department of Physics and Astronomy\\
University of Rochester\\
Rochester, NY 14627}

\date{\today}
\maketitle
\begin{abstract}
The dynamics of the random-phase sine-Gordon model, which
describes $2D$ vortex-glass
arrays and crystalline surfaces on disordered substrates,
is investigated using the self-consistent Hartree approximation. The
fluctuation-dissipation theorem is violated below the critical temperature
$T_c$ for large time $t>t^{\ast}$ where $t^{\ast}$ diverges in the
thermodynamic limit. While above $T_c$ the averaged autocorrelation function
diverges as $T\ln(t)$, for $T<T_c$
it approaches a finite value $q^{\ast} \sim 1/(T_c-T)$ as
$q(t) = q^{\ast} - c\,(t/t^{\ast})^{-\nu}$ (for $t\rightarrow t^{\ast}$)
where $\nu$ is a temperature-dependent exponent. On larger
time scales $t > t^{\ast}$ the dynamics becomes non-ergodic.
The static correlations behave as $\sim T \ln|\vec{x}|$ for $T>T_c$ and
for $T<T_c$ when $x<\xi^{\ast}$  with
$\xi^{\ast}\sim \exp\left\{A/(T_c-T)\right\}$.
For scales $x > \xi^{\ast}$, they behave as $\sim m^{-1}T\ln|\vec{x}|$
where $m\approx T/T_c$ near $T_c$, in general agreement with the variational
replica-symmetry breaking approach and with recent simulations of the
disordered-substrate surface. For strong-coupling the transition becomes
first-order.

\end{abstract}
\narrowtext

   Much attention has been given recently to the $2D$ sine-Gordon model
\cite{Cardy_Ostlund,Gold_Houg_Sch,Vill_Fer,Toner_Vic,MPFisher,Tsai_Shapir,Korshunov,Giam_LeDou,Tsvelik,Balents_Kardar,Hwa_DFisher,Batrouni_Hwa,Cule_Shapir,LeDou_Giam,Nat_Lyuk,Toner}
with random phases which
describes two very interesting disordered systems: $(i)$ An array of flux-lines
as obtained when a magnetic field is applied parallel to a type $II$
superconducting film \cite{MPFisher}, and $(ii)$ A $3D$ crystalline surface
with
a disordered substrate \cite{Tsai_Shapir}. It is also equivalent to a
vortex-free $XY$ model with a random field.

   The intensive theoretical attention is due to diverging results obtained
by the renormalization group (RG)
\cite{Cardy_Ostlund,Gold_Houg_Sch,Vill_Fer,Toner_Vic},
the Gaussian variational (GV)
approximation \cite{Korshunov,Giam_LeDou}, $n\rightarrow 0$ Bethe ansatz
\cite{Tsvelik,Balents_Kardar}, and by two extensive numerical studies
\cite{Batrouni_Hwa,Cule_Shapir}. The Hamiltonian of the random sine-Gordon
model (RSGN) is:

\begin{equation}
 H = \int d\vec{x} \left\{ \frac{\kappa}{2}
     \left[ \nabla h(\vec{x},t) \right]^2
            - g \cos \left( \gamma \left[ h(\vec{x},t) - d(\vec{x})
     \right] \right) \right\}.
\label{E1}
\end{equation}
\noindent
The coefficient $\kappa$ is the stiffness, $g$ is the coupling constant,
$\gamma$ determines the periodicity (we choose $\gamma=2\pi$), and
$d(\vec{x})$ is a random variable such that $\gamma d(\vec{x})$ is a
 random phase uniformly distributed in $(0,2\pi]$.
Random phases for different $\vec{x}$ are uncorrelated
(short-distance correlations will not affect the conclusions).
The variable $h(\vec{x},t)$ denotes the vortex-line displacement from a
periodic lattice or the height of the surface.

 The controversial issues concern the properties of the low-temperature phase.
The quantity of interest is the correlation function:

\begin{equation}
 C(\vec{x}-\vec{x}',t-t') = \left< \left[ h(\vec{x},t) - h(\vec{x}',t')
                            \right]^2 \right>,
\label{E2}
\end{equation}
\noindent
where $\left< \cdots \right>$ denotes averaging over the thermal noise and the
quenched disorder.

   In the static limit, RG predicts
\cite{Cardy_Ostlund,Gold_Houg_Sch,Vill_Fer,Toner_Vic}  that
$C(\vec{x}-\vec{x}')$
changes behavior from $(T/\pi\kappa) \ln|\vec{x}-\vec{x}'|$
for $ T>T_c$, to a behavior $\sim \epsilon^2 (\ln|\vec{x}-\vec{x}'|)^2$
for $T < T_c$ where
$T_c=\kappa/\pi$ is the critical temperature and $\epsilon = |1-T/T_c|$.
The GV approach yields \cite{Korshunov,Giam_LeDou} a one-step replica
symmetry breaking (RSB) which gives correlations which remain logarithmic but
with a different coefficient $S(T)\ln|\vec{x}-\vec{x}'|$ for $T<T_c$.
One study \cite{Giam_LeDou} finds $S(T) = T_c/\pi\kappa$ while
the other \cite{Korshunov} finds a coefficient which {\it increases}
from this value for temperature below $T_c$.

  A potential resolution of the contradiction between the RG and the
GV results was suggested recently by Le Doussal and Giamarchi
\cite{LeDou_Giam}.
They have shown that at the replica-symmetric RG fixed-line for $T<T_c$
RSB is a relevant perturbation. This suggest the possibility of a
spontaneous replica-symmetry breaking.

  Numerical simulations of the weak coupling RSGM show no sign of
a transition whatsoever in the equilibrium properties \cite{Batrouni_Hwa}.
Our simulations of the random-substrate model, which is better
described by the strong-coupling RSGM, exhibit a continuous transition
at the analytically predicted temperature \cite{Cule_Shapir}.
The behavior of the correlations is consistent with the predictions
of the GV theory with one-step RSB.

   For the dynamical behavior the only analytic predictions are from
heuristic arguments \cite{Nat_Lyuk,Toner} and from RG calculations
\cite{Tsai_Shapir}. The RG predicts anomalous dynamics for $T<T_c$ which
is manifested by a temperature-dependent dynamic exponent such that
$z = 2(1 + 1.78\epsilon)$ and a non-linear response to a small driving
force $F$. The average velocity is predicted to be proportional to $F^{z/2}$
for small $F$. The simulations \cite{Batrouni_Hwa} vindicate qualitatively
this latter prediction although the coefficient of $\epsilon$ is much smaller
than that predicted by dynamic RG.

  In view of these results we study here a non-perturbative self-consistent
approach to the dynamics of the system as a complement to the RG calculations.
The general approach we take is based on that introduced by Sompolinsky
and Zippelius (SZ) \cite{Som_Zip} for the Sherrington-Kirkpatrick (SK) model.
This approach has been extended by Crisanti, Horner, and Sommers for the
spherical $p$-spin interacting spin-glass model \cite{CHS} and
generalized by Kinzelbach and Horner (KH) \cite{Kin_Hor} for
fluctuating manifolds in disordered media.

    We begin by the RSGM Hamiltonian (\ref{E1}) with random-quenched phase
with zero mean and correlations:

\begin{equation}
 \left< e^{\imath \gamma d(\vec{x})} e^{-\imath\gamma d(\vec{x}')} \right>
     = \delta(\vec{x}-\vec{x}').
\label{E3}
\end{equation}
\noindent
The stochastic equation of motion obeyed by $h(\vec{x},t)$ within
purely relaxational dynamics is:

\begin{equation}
\frac{1}{\Gamma_0}\frac{\partial h(\vec{x},t)}{\partial t} =
   -\beta \frac{\delta H[h]}{\delta h(\vec{x},t)} + \zeta(\vec{x},t).
\label{E4}
\end{equation}
\noindent
The Gaussian white noise $\zeta(\vec{x},t)$ obeys
$\left< \zeta(\vec{x},t)\zeta(\vec{x}',t') \right> =
(2/\Gamma_0)\delta(\vec{x}-\vec{x}')\delta(t-t')$. $\Gamma_0$ is a microscopic
kinetic coefficient and $\beta = 1/T$.
The next step is to obtain in a standard way the
Martin-Siggia-Rose (MSR) generating functional \cite{MSR} which may
directly be averaged over the noise $\zeta$ and the quenched disorder
$d(\vec{x})$. The action so obtained reads:

\begin{eqnarray}
 S &=& \int d\vec{x} \int dt \left\{ -\Gamma_{0}^{2}
  \left[ \imath \hat{h}(\vec{x},t) \right]^2  +
\imath \hat{h}(\vec{x},t) \left[ \partial_t -\Gamma_0\kappa\beta
\nabla^2\right]
h(\vec{x},t) \right\} \nonumber \\
   &-& \frac{1}{4}\beta^2\Gamma_{0}^{2}\gamma^2 g^2 \int d \vec{x} \int dt dt'
  \;\imath \hat{h}(\vec{x},t)\imath\hat{h}(\vec{x},t')
   \cos\left[ \gamma\left( h(\vec{x},t) - h(\vec{x},t') \right) \right].
\label{E5}
\end{eqnarray}
\noindent

In the self-consistent Hartree approximation all terms which contains more
than two fields $h$ or $\hat{h}$ are written in all possible ways
as products of such pairs and all pairs but one are replaced by
their expectation values. The action then is quadratic in $h$ and $\hat{h}$
with coefficients which are only functions of response and correlation
functions, defined as:

\begin{eqnarray}
r(\vec{x}-\vec{x}',t-t') &=& \left< h(\vec{x},t)
 \imath \hat{h}(\vec{x}',t') \right>, \hspace{.7in} t> t'
\\
 q(\vec{x}-\vec{x}',t-t') &=& \left< h(\vec{x},t) h(\vec{x},t) \right> -
  \left< h(\vec{x},t)h(\vec{x}',t') \right>.
\label{E6-7}
\end{eqnarray}
\noindent
Their Fourier coefficients are denoted by $r(\vec{k},t)$ and $q(\vec{k},t)$.
The effective action obtained in this way is:

\begin{eqnarray}
 S_{eff} &=& \int d\vec{x} \int dt \left\{ -\Gamma_{0}^{2}
  \left[ \imath \hat{h}(\vec{x},t) \right]^2  +
\imath \hat{h}(\vec{x},t) \left[ \partial_t -\Gamma_0\kappa\beta \nabla^2
  + \mu(0)\right]
h(\vec{x},t) \right\} \nonumber \\
   &-& \int d \vec{x} \int dt dt'
   \left\{
 \frac{1}{2}\omega(t-t') \imath \hat{h}(\vec{x},t)\;\imath\hat{h}(\vec{x},t')
   + \lambda(t-t')\imath\hat{h}(\vec{x},t)h(\vec{x},t') \right\}.
\label{E8}
\end{eqnarray}
\noindent
where we have defined $(\Gamma_0=1, \gamma=2\pi)$ :

\begin{equation}
\omega(t)=\frac{1}{2}\beta^2\Gamma_{0}^{2}\gamma^2 g^2
  \left< \cos \left[ 2\pi \left( h(\vec{x},t) - h(\vec{x},t') \right) \right]
  \right> = \frac{2 g^2 \pi^2}{T^2}e^{-4\pi^2 q(t)},
\label{E9}
\end{equation}
\noindent
$\lambda(t) = 4\pi^2 \omega(t) r(t)$ and
$\mu(t) = \int_{t}^{\infty} dt' \lambda(t')$. The functions $\omega$, $\lambda$
and $\mu$ depend on $t$ only through $q(t) = q(\vec{0},t)$ and
$r(t) = r(\vec{0},t)$.

 The next step is to obtain the equations of motion. By standard techniques
explained in details in Ref. \cite{Kin_Hor}, in momentum space they read:

\begin{equation}
\left\{ \partial_t + \beta\kappa k^2 +  \mu(o) \right\} r(\vec{k},t)  -
  \int_{0}^{t} dt' \lambda(t-t')r(\vec{k},t') = 0
\label{E10}
\end{equation}
\noindent
with initial condition $r(\vec{k},0^{+}) = 1$, $r(\vec{k},t) = 0$ for
$t \leq 0$, and:

\begin{equation}
\left\{ \partial_t + \beta\kappa k^2 +  \mu(o) \right\} q(\vec{k},t)  -
  \int_{0}^{t} dt' \lambda(t-t')q(\vec{k},t') =
  1 - \left\{ I(\vec{k},t) - I(\vec{k},0) \right\}
\label{E11}
\end{equation}
\noindent
 with $q(\vec{k},0) = 0$ and $I(\vec{k},t) = \int_{0}^{\infty}dt' \left\{
\omega(t+t')r(\vec{k},t') -\lambda (t+t')q(\vec{k},t') \right\}$.
If fluctuation-dissipation theorem (FDT)
holds, the $I(\vec{k},t) -I(\vec{k},0) = 0$ and equations for $r$ and $q$
become equivalent as they should. The validity of the FDT implies that for
long time:

\begin{equation}
\left[ \beta\kappa k^2 +  \mu(t) \right] q(\vec{k},t)  \leq 1.
\label{E12}
\end{equation}
\noindent
This condition may be cast as:

\begin{equation}
 q(t) \leq \frac{T}{4\pi\kappa}\ln \left( 1 + \frac{\kappa\pi^2}{\mu(t) T}
       \right).
\label{E13}
\end{equation}
\noindent
Following KH \cite{Kin_Hor} we define the function:

\begin{equation}
\Delta(q) =\frac{\kappa\pi^2}{T}\left[ \frac{1}{e^{4\pi\kappa q/T} -1} -
       \frac{2g^2}{\kappa T} e^{-4\pi^2 q} \right]
\label{E14}
\end{equation}
\noindent
in terms of which the FDT dynamics holds as long as $\Delta(q) \geq 0$.
However, this condition will be violated at low enough temperatures.
For $g > g_{tr} = \kappa/\sqrt{2\pi}$ the transition is ``first order''
in the sense that as $T$ is lowered $\Delta(q)$ vanishes at a finite $q$.
We shall postpone the
discussion of the first-order regime and the tricritical behavior which
takes place at $g_{tr}$ to a future, longer, publication. Here we concentrate
on the regime $g<g_{tr}$ for which the transition is continuous.
In this regime $\Delta(q)$ becomes negative at the $g$-independent
critical temperature $T_c = \kappa/\pi$
(as found by other analytic approaches). In this regime, and close to $T_c$,
$\Delta(q)\leq0$ for $q\geq q_1=(1/4\pi^2\epsilon)|\ln(2g^2/\kappa T)|$.

  The solution proposed assumes the existance of a sharp separation of
time scales between the ergodic behavior (i.e., relaxation within pure
states) for $t \ll t^{\ast}$ and the non-ergodic behavior for $t\gg t^{\ast}$
which includes ``tunneling'' between different pure-states. In this
latter regime  $\lambda(t) < |\partial_t \omega|$ and the integral
$\int_{t^\ast}^{\infty} dt' \lambda(t') = \mu^{\ast} > 0$.
Therefore for $t \ll t^{\ast}$, $\tilde{\Delta}(q) = \Delta(q) +\mu^{\ast}$
and the instability is avoided.
Within the so-called ``quasi-FDT'' (QFDT) approach \cite{CHS,Kin_Hor,Horner}
it is assumed that $\mu^{\ast}$ is the minimal one necessary to make
$\tilde{\Delta}$ non-negative.
This choice implies $\tilde{\Delta}'(q^{\ast})= 0$,
where $q^{\ast}$ is derived from the ergodic stationary solution
as $t\rightarrow t^{\ast}$:
$q^{\ast}(\vec{k}) = 1/(\beta\kappa k^2 + \mu^{\ast})$ and
therefore $q^{\ast}= (T/4\pi\kappa)\ln(1+ \kappa \pi^2/\mu^{\ast}T)$.

At long-time $t\gg t^{\ast}$ the time variable rescales to $\tau=t/t^{\ast}$.
We look for the solutions for the response and correlation
functions which have the
scaling forms $q(\vec{k},t) = \tilde{q}(\vec{k},t/t^{\ast})$ and
$r(\vec{k},t) = (1/t^{\ast}) \tilde{r}(\vec{k},t/t^{\ast})$ with similar
expressions for $\omega(t)$ and $\lambda(t)$. In the
limit $t \rightarrow \infty$ the equations for $\tilde{q}$ and $\tilde{r}$
have solution of the form:

\begin{equation}
m\partial_{\tau} \tilde{q}(\vec{k},\tau) =\tilde{r}(\vec{k},\tau),
 \mbox{\hspace{.5in}} t> t^{\ast}
\label{E15}
\end{equation}
\noindent
which implies:

\begin{equation}
m\partial_{\tau} \tilde{\omega}(\tau) =-\tilde{\lambda}(\tau),
 \mbox{\hspace{.5in}} t> t^{\ast}.
\label{E16}
\end{equation}
\noindent
Then the equations for $\tilde{q}$ and $\tilde{r}$ are equivalent and
hence quasi-FDT holds. Equation (\ref{E16}) also
determines $m$ since its integration leads to:

\begin{equation}
m= \frac{\mu^{\ast}}{\omega^{\ast}}\;,
\label{E17}
\end{equation}
\noindent
where $\omega^{\ast} = \omega(q^{\ast})$. Using the equation for $q^{\ast}$,
$(\Delta'(q^{\ast}) = 0 )$, we find

\begin{equation}
m(T)= \frac{\pi T}{\kappa} \left[ 1 - e^{-4\pi\kappa q^{\ast}/T} \right].
\label{E18}
\end{equation}
\noindent

 We note that as $t \rightarrow t^{\ast}$ the solution for
$t>t^{\ast}$ has to merge with the
one below $t^{\ast}$.
This matching condition  yields the equation:

\begin{equation}
1 = 4\pi^2 \omega^{\ast} \int \frac{d\vec{k}}{(2\pi)^2}
   \frac{1}{\left( \beta\kappa k^2 + \mu^{\ast} \right) ^2 }
\label{E19}
\end{equation}
\noindent
which is exactly the condition $\Delta'(q^{\ast}) = 0$, or in the replica
language it is the condition on $\mu^{\ast}$ for the eigenvalue of the replicon
mode to change from a negative value to zero (see Eq. (5) in Ref.
\cite{Giam_LeDou}).

For $t \gg t^{\ast} \rightarrow \infty$, the stationary value of
$q(\vec{k})$ is now given by:

\begin{equation}
q_{\infty}(\vec{k}) = \frac{T}{\kappa k^2}
  \left[ 1 + (1-m) \omega^{\ast} q^{\ast} (\vec{k}) \right]
\label{E20}
\end{equation}
\noindent
which leads to:

\begin{equation}
q_{\infty}(\vec{k}) = \frac{T}{m\kappa k^2}
 - \frac{1-m}{m} \frac{T}{\kappa(k^2 + \mu^{\ast} T/\kappa )}.
\label{E21}
\end{equation}
\noindent

  In the regime $T<T_c$ the equal-time correlations are:

\begin{equation}
  C(\vec{x}) = \left\{
            \begin{array}{lr}
            \frac{T}{\pi\kappa}\ln|\vec{x}| & $if$\;\;  x\ll \xi^{\ast} \\
           \frac{T_c}{\pi\kappa}\left(1 - e^{-4\pi\kappa
q^{\ast}/T}\right)^{-1}
            \ln|\vec{x}| &\hspace{.5in} $if$\;\; x\gg \xi^{\ast}
            \end{array}
            \right.
 \label{E22}
\end{equation}
\noindent
where $\xi^{\ast} =\sqrt{\kappa/T\mu^{\ast}} \sim \exp(A/\epsilon) $.
As $T_c$ is approached from below the slope of the logarithm of
the asymptotic behavior is given by:

\begin{equation}
  \frac{1}{\pi^2}\left( 1  +\left[ 2\pi g^2/\kappa^2\right]^{1/\epsilon}
     \right)
\label{E23}
\end{equation}
\noindent
which is consistent with the behavior found by Korshunov \cite{Korshunov}.
Fig. \ref{F1} shows the slope $S(T)$ for different values of $g$.
Near $T_c$ the behavior is also consistent with Ref. \cite{Giam_LeDou}
and with the numerical results we obtained simulating the random-substrate
model
\cite{Cule_Shapir}.

  The dynamics within each ergodic component for time $t<t^{\ast}$ for
$T<T_c$ is studied by assuming a solution of the form

\begin{equation}
q(t) = q^{\ast} - \delta q(t).
\label{E24}
\end{equation}
\noindent
Expanding in $\delta q(t)$ we find a solution of the form
$\delta q(t) \sim t^{-\nu}$ where $\nu$ is a
temperature dependent exponent determined from the solution of the
equation \cite{CHS,Kin_Hor} :

\begin{equation}
\frac{\Gamma^2(1-\nu)}{\Gamma(1-2\nu)} = m(T),
\label{E25}
\end{equation}
\noindent
where $\Gamma$ is the gamma function.  The plot of $\nu(T)$ is given in
Fig. \ref{F2}.

So, while above $T_c$, $q(t)$ diverges as $(T/4\pi\kappa)\ln(t)$, below $T_c$
it approaches a finite value as $t\rightarrow t^{\ast}$
(and $t^{\ast} \rightarrow \infty$ as $L \rightarrow \infty$ ),
$q(t) = q^{\ast} - c\; t^{-\nu}$. As $T \rightarrow T_{c}^{-}$,
$\nu \rightarrow 0$ and the approximation (\ref{E24}) should be handled
carefully.  The asymptotic behavior becomes logarithmically slow. In this limit
the value of $q^{\ast}$ diverges like $q^{\ast} \sim  1/(T_c-T)$.

If we keep the size $L$ dependence explicitly and take the limit
$L\rightarrow \infty$ after $t\rightarrow \infty$, the behavior of $q(t)$
will be modified as $T_c$ is approached from above.
This will be
studied in a future paper, where we shall also address the
first order transition
which occurs in the strong coupling regime. In this regime the dynamic critical
temperature is g-dependent and larger than the static one. Such a behavior
was found in
other systems with first-order transition \cite{CHS,Kin_Hor} and it
signals the existence of many metastable states \cite{Mar_Par_Rit}.

 To summarize, in the second-order regime the self-consistent
dynamic approach yields the following results: In equlibrium it essentially
reproduces
the results of the GV approach with the one-step replica
symmetry breaking. In the dynamics this is refelected by the
breaking of the FDT below $T_c$ on large time scales $t\geq t^{\ast}$.
Phase space is broken into separate components and the divergence of
their barriers leads to the divergence of $t^{\ast}$ in the thermodynamic
limit.
Within each component the autocorrelations $q(t)$ approach a finite
value $q^{\ast}$. At short time $q(t) \sim \ln(t)$ and $q(t) \sim q^{\ast} -
c\:t^{-\nu}$ as $t\rightarrow t^{\ast}$. $q^{\ast}$ diverges and
$\nu \rightarrow 0$ as $T \rightarrow T_{c}^{-}$.   This behavior is
obtained based on the self-consistent quasi-FDT which holds
for $t>t^{\ast}$. The analysis applies if $t\rightarrow \infty$
before $L\rightarrow \infty$, where $L$ is the system size. If the
order of limits is reversed the system never reaches equilibrium below $T_c$.
This will entail a non-stationary behavior with aging phenomena
\cite{Cug_Kur}.

  This research was supported in part by the National Science Foundation
under Grant No. PHY89-04035. Acknowledgment is also made to the donors
of The Petroleum Research Fund, administrated by ACS, for partial
support of this research.

\begin{figure}
\caption{
 The slope of the $\ln|\vec{x}|$ term in
 Eq. (\protect\ref{E22}) versus temperature for different coupling
 constants $g < g_{tr}$ and $\kappa=1$.
}
\label{F1}
\end{figure}

\begin{figure}
\caption{
 The temperature dependence of the critical exponent $\nu$ for different
 coupling constants $g$. As $g$ decreases they approach the solution of the
 Eq. (\protect\ref{E24})  with $m =\pi T/\kappa$.
}
\label{F2}
\end{figure}

\end{document}